\documentclass[%
 reprint,
 amsmath,amssymb,
 aps,
nofootinbib
]{revtex4-2}

\pdfoutput=1 
\usepackage{graphicx}
\usepackage{dcolumn}
\usepackage{bm}
\usepackage{amssymb}
\usepackage{amsmath}
\usepackage{commath}
\usepackage{graphicx,bm}
\usepackage{verbatim}
\usepackage[caption=false]{subfig}

\usepackage[colorlinks=true,linkcolor=blue,citecolor=blue,allcolors=blue]{hyperref}%
\usepackage{soul}

\usepackage{color}
    \definecolor{darkgreen}{rgb}{0,0.5,0}
    \definecolor{darkred}{rgb}{0.5,0,0}
    \definecolor{darkblue}{rgb}{0,0,0.6}
    \definecolor{purple}{rgb}{0.4,.2,0.7}

\begin{document}

\title{The Low Energy Limit of BFSS Quantum Mechanics}
\author{\'Oscar~J.~C.~Dias,}
\affiliation{STAG research centre and Mathematical Sciences,  University of Southampton, University Road Southampton SO17 1BJ, U.K.}
\email{ojcd1r13@soton.ac.uk}
\author{Jorge~E.~Santos}
\affiliation{DAMTP, Centre for Mathematical Sciences, University of Cambridge, Wilberforce Road, Cambridge CB3 0WA, U.K.}
\email{jss55@cam.ac.uk}

\date{\today}

\begin{abstract}
We investigate the low-energy regime of BFSS quantum mechanics using its holographic dual. We identify three distinct thermodynamic phases (black holes) and analyze their thermodynamic properties extensively, including phase transitions amongst the several phases. While the properties of the canonical ensemble aligns with existing conjectures on BFSS thermodynamics, we uncover intriguing and unexpected behavior in the microcanonical ensemble. Specifically, for sufficiently low energies, we observe the dominance of the localized  phase. Surprisingly, we also identify an energy range where the non-uniform phase becomes dominant. The transition between these phases is mediated by a Kol-type topology-changing phenomenon.
\end{abstract}

\maketitle

\paragraph*{\bf Introduction.}
The holographic principle, also known as gauge/gravity duality, connects certain quantum field theories (QFT) to higher-dimensional quantum gravity, and it is foundational in modern high-energy physics. In this context, finite temperature phases in the QFT correspond one-to-one with black hole phases in the gravitational theory, including their thermodynamic properties.
Maldacena's original duality, proposed in \cite{Maldacena:1997re}, relates ten-dimensional type IIB supergravity on AdS$_5 \times S^5$ to four-dimensional $\mathcal{N}=4$ supersymmetric Yang-Mills (SYM) with large $N$ and strong 't Hooft coupling. Despite substantial supporting evidence, no formal proof of this duality exists, as the gravitational descriptions apply only when the gauge theory duals are strongly coupled \cite{Maldacena:1997re, Witten:1998qj, Aharony:1999ti}. Additionally, $\mathcal{N}=4$ SYM, though highly symmetric, is impractical to simulate with standard lattice Monte Carlo methods. Its four-dimensional nature introduces computational complexity, and its supersymmetric  ultraviolet (UV) regime includes many fermionic degrees of freedom, complicating standard path integral formulations due to the sign problem \cite{Loh:1990zz} (recently, there has been however significant progress dealing with these obstacles~\cite{Bergner:2021goh,Catterall:2023tmr}).
 
Shortly after \cite{Maldacena:1997re},  Ref.~\cite{Itzhaki:1998dd} proposed a network of  additional holographic dualities pairing maximally supersymmetric theories with their gravitational duals (see also~\cite{Polchinski:1999br,Boonstra:1998mp,Kanitscheider:2008kd}). These were identified by studying D-branes in string theory and decoupling the brane worldvolume theory from gravity. This letter focuses on the simplest duality of  \cite{Itzhaki:1998dd}, namely, between the low-energy limit of D$0$-branes and maximally supersymmetric quantum mechanics (BFSS) \cite{Banks:1996vh}. Recent years have seen the development of novel techniques for analyzing the gauge theory side of this duality at strong coupling using computer lattice simulations \cite{Hanada:2007ti, Catterall:2007fp, Anagnostopoulos:2007fw, Catterall:2008yz, Hanada:2008gy, Hanada:2008ez, Catterall:2009xn, Hanada:2009ne, Hanada:2013rga,Hanada:2016pwv,Bergner:2021goh,Hanada:2023rlk}. Our study will offer key insights into the gravitational aspect of the duality, crucial for understanding BFSS theory in its strongly coupled regime and enabling precise holography matchings.

BFSS is a gauge theory that includes finite $N \times N$ traceless Hermitian bosonic degrees of freedom $X^j$ and fermionic degrees of freedom $\theta^{\alpha}$, transforming as spinors of $SO(9)$. The action is:
\begin{multline}
S_{\rm BFSS}=\frac{N}{2\lambda}\int {\rm d}t\;{\rm Tr}\Big\{(D_t X^j)^2+\theta^{\alpha}D_t \theta^{\alpha}
\\
+\frac{1}{2}\left[X^j,X^k\right]^2+{\rm i}\theta^{\alpha}\gamma^j_{\alpha \beta}\left[\theta^{\beta},X^j\right]\Big\}\,,
\end{multline}
where $\lambda\equiv g_{\rm YM}^2N$ is the t'Hooft coupling, $g_{\rm YM}$ is the gauge coupling, and $D_t$ is the gauge covariant derivative with respect to time $t$. Summation over spatial indices $i, j = 1, \ldots, 9$  and spinor indices $\alpha, \beta = 1, \ldots, 16$ is implicit. $\lambda$ has units of energy cubed so, at finite temperature $\mathcal{T}_H$, the system's thermodynamics depends on the dimensionless parameters $\tau \equiv \mathcal{T}_H/\lambda^{1/3}$ and $N$.

According to \cite{Itzhaki:1998dd}, a dual gravitational description involving eleven-dimensional supergravity is expected for
\begin{equation}
N^{-5/6} \ll \tau \ll 1,
\end{equation}
where the lower bound ensures black holes are larger than the eleven-dimensional Planck length and the upper bound ensures small curvatures in string units. When a classical description is adequate, the action for this supergravity is:
\begin{equation}
S_{\rm 11D}=\frac{1}{16\pi G_{11}}\int \!\!\left(\eta\,R+\frac{1}{2}\mathrm{d}C \!\wedge \star \mathrm{d}C-\frac{1}{6}C \!\wedge\! {\rm d}C \!\wedge\! {\rm d}C\right)\!,
\label{eq:11D}
\end{equation}
where $\eta$ is the spacetime volume form, $R$ is the Ricci scalar, $C$ is a 3-form gauge potential, $\star$ is the standard 11-dimensional Hodge dual operation, and $G_{11}$ is the 11-dimensional Newton constant. In this letter, we will focus on geometries with $C=0$.

Following \cite{Itzhaki:1998dd}, we search for solutions to the equations of motion derived from (\ref{eq:11D}) that asymptotically approach, at large $r$, the metric:
\begin{equation}
{\mathrm d}s^2= -\frac{r^7}{\mathcal{R}^7}{\rm d}t^2+{\rm d}r^2+\frac{\mathcal{R}^7}{r^7}\left(\mathrm{d}z+\frac{r^7}{\mathcal{R}^7}{\rm d}t\right)^2+r^2\mathrm{d}\Omega_8^2\,,
\label{eq:asym}
\end{equation}
where $\mathrm{d}\Omega_8^2$ is the line element of a unit radius sphere $S^8$ and $z\sim z+L_{11}$ with $L_{11}=2\pi g_s\ell_s$ parametrising the M-theory circle ($\ell_s$, $g_s$ are the string length and string coupling constant, respectively). The map between BFSS and 11-dimensional supergravity is:
\begin{equation}
g_s=\frac{4\pi^2\ell_s^3}{N}\lambda\,,\quad G_{11}=16\pi^7g_s^3\ell_s^9\,,
\quad \left(\frac{\mathcal{R}}{\ell_s}\right)^7=60\pi^3g_s N.
\end{equation}


For sufficiently high temperatures, namely when $N^{-5/9}\lesssim\tau\ll1$, \cite{Itzhaki:1998dd} conjectured that the dominant phase retains translational invariance in $z$, has horizon topology $S^1\times S^8$, and is simply given by
\begin{subequations}\label{eq:larger}
\begin{equation}
{\mathrm d}s^2= -\frac{r^7}{\mathcal{R}^7}f{\rm d}t^2+\frac{{\rm d}r^2}{f}+\frac{\mathcal{R}^7}{r^7}\left[\mathrm{d}z+\frac{r^7}{\mathcal{R}^7}f(r){\rm d}t\right]^2+r^2\mathrm{d}\Omega_8^2\,,
\end{equation}
with
\begin{equation}
f(r)=1-\frac{r_0^7}{r^7}\quad\text{where}\quad \left(\frac{r_0}{\ell_s}\right)^5=\frac{120\pi^2}{49}\left(2\pi g_s N \right)^{5/3}\tau^2.
\end{equation}
\end{subequations}
This solution describes a stack of uniform D0 branes at a finite temperature $\tau$ extended along the M-theory circle.

However, for sufficiently low temperatures, specifically $N^{-5/6}\ll\tau\lesssim N^{-5/9}$, \cite{Itzhaki:1998dd} conjectured that the relevant phase should be a localized black hole on the M-theory circle. In this letter, we will show that while this expectation is met in the canonical ensemble, it surprisingly fails in the microcanonical ensemble. Indeed, we will present compelling numerical evidence for the existence of a dominant phase, within a specific range of energies, that breaks translational invariance in $z$ while maintaining a horizon topology of $S^1\times S^8$. Morover, in the canonical ensemble, we will determine the precise temperature at which the first-order transition occurs between localised black holes and the translationally preserving phase.

We will describe the construction of black hole phases in supergravity, map them to BFSS thermodynamic phases, and present BFSS phase diagrams in both canonical and microcanonical ensembles.
\paragraph*{\bf Constructing the black hole phases.} 
The black holes/ strings we aim to construct can be derived from Ricci-flat, \emph{static} configurations with Kaluza-Klein asymptotics $\mathbb{R}^{1,9}\times S^1_L$. We will detail the construction of these configurations and then present the mapping that transforms such solutions to those approaching \eqref{eq:asym}
at large $r$.

We use the DeTurck trick, introduced in \cite{Headrick:2009pv} and  reviewed in \cite{Dias:2015nua,Wiseman:2011by}. The method starts by considering the so called Einstein-DeTurck equation
\begin{equation}
R_{ab}-\nabla_{(a}\xi_{b)}=0
\label{eq:ede}
\end{equation}
with $\xi$ being the DeTurck vector, given by
\begin{equation}\label{eq:ede2}
\xi^{a} = g^{cd} \left[\Gamma^{a}_{cd}(g)-\Gamma^{a}_{cd}(\bar{g})\right]
\end{equation}
where $\Gamma^{a}_{cd}(\mathfrak{g})$ is the Christoffel connection associated to a metric $\mathfrak{g}$, and $\bar{g}$ is a reference metric which we are free to choose. Solutions to the Einstein-DeTurck equation are generally not always Ricci-flat. However, if we focus on static configurations that admit a regular Euclidean section, and have Kaluza-Klein asymptotics, it has been shown in \cite{Figueras:2011va} that solutions with $\xi\neq0$ do not exist. As such, for such spacetimes, we expect solutions of the Einstein-DeTurck equation to coincide with Ricci-flat solutions. Solving (\ref{eq:ede}) instead of $R_{ab}=0$ offers significant advantages. Once a reference metric $\bar{g}$ is chosen, the Einstein-DeTurck equation forms a well-posed elliptic system \cite{Figueras:2011va}. Solutions must have $\xi^a=0$, implying $\Box x^a=g^{cd}\Gamma^{a}_{cd}(\bar{g})$. Thus, these solutions are naturally expressed in generalized harmonic coordinates \cite{cmp/1104114004}, with the gauge source term fully controlled by $\bar{g}$. A good gauge choice involves choosing $\bar{g}$ judiciously. The choices of $\bar{g}$ across the solutions described in this letter are presented in the Supplemental Material.

\begin{figure}[tb]
    \centering    \includegraphics[width=0.45\textwidth]{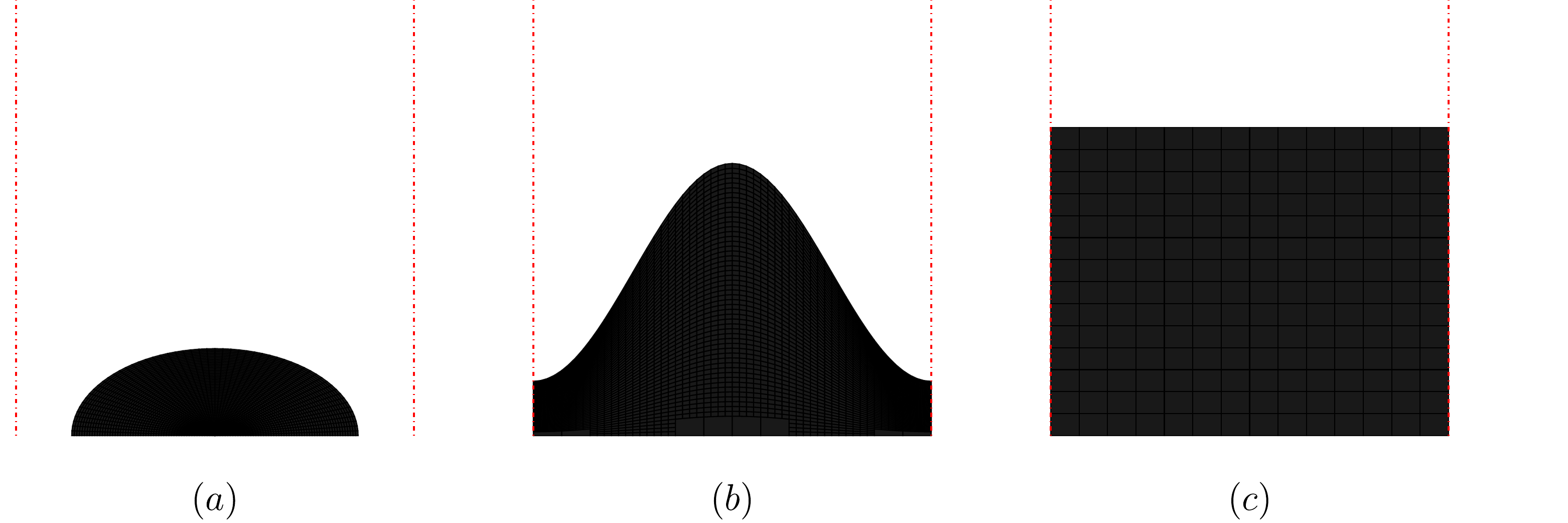}
    \caption{A schematic drawing of the transverse horizon radius for different black hole phases with standard Kaluza-Klein asymptotics: (a) the localized black hole phase, (b) the non-uniform string phase, and (c) the uniform string phase.}
    \label{fig:sketch}
\end{figure}

Since the solutions we aim to construct first are static, they possess a hypersurface orthogonal Killing vector field $k$. By introducing a time coordinate $T$ that parametrizes the integral curves of $k$, we have $k=\partial/\partial T$. Additionally, these solutions are spherically symmetric with an $SO(9)$ isometry group. The most general line element compatible with such symmetries reads
\begin{align}
\mathrm{d}s^2 &=-Q_1(X,Z)\,\mathrm{d}T^2+\frac{4\,Q_2(X,Z){\rm d}X^2}{(1-X^2)^4}
\\
& \hspace{-0.4cm}+L^2Q_4(X,Z)\left[\mathrm{d}Z+Q_3(X,Z){\rm d}X\right]^2+\frac{Q_5(X,Z)}{(1-X^2)^2}\mathrm{d}\Omega_8^2\,, \nonumber
\end{align}
where $Z\sim Z+1$. The factors of $(1-X^2)$ were chosen so that near the asymptotic boundary, $r\sim \frac{1}{1-X^2}$, implying $X\to1^-$ as $r\to+\infty$. Near $X=1$, we demand $Q_1=Q_4=1$, $Q_2=Q_5=K^2$, with $K>0$ constant, and $Q_3=0$ so that the spacetime approaches
\begin{equation}
\mathrm{d}s^2=-\mathrm{d}T^2+{\rm d}r^2+r^2\mathrm{d}\Omega_8^2+L^2\mathrm{d}Z^2\,,
\end{equation}
at large $r$, showing that it exhibits the desired asymptotic structure. All the functions $Q_i$ are periodic with respect $Z$, but as $X$ becomes small these functions behave differently across the different black hole phases. In Fig.~\ref{fig:sketch} we sketch the expected behaviour for the different black hole topologies.

There are three phases of interest. In the uniform phase (right-most panel of Fig.~\ref{fig:sketch}) translational invariant along the $Z$ direction is preserved. In this phase the functions $Q_i$ can be found analytically:
\begin{multline}
Q_1(X,Z)=X^2\sum_{i=0}^6(1-X^2)^i\,,\quad Q_2(X,Z)=\frac{X^2r_0^2}{Q_1(X,Z)}\,,
\\
Q_3(X,Z)=0\,,\quad Q_4(X,Z)=1\quad\text{and}\quad Q_5(X,Z)=r_0^2\,,
\label{eq:uni}
\end{multline}
with $r_0/L$ parametrising the thickness of the black string in terms of the length $L$ of the Kaluza-Klein circle and with the black hole event horizon being located at $X=0$. This solution is simply ${\rm Schw}_{10D}\times S^1$, with ${\rm Schw}_{10D}$ being a ten-dimensional Schwarzschild black hole. Spatial cross sections of the event horizon have topology $S^1\times S^8$. Hereafter, phases with this topology will be referred to as black strings.

It has long been known \cite{Sorkin:2004qq} that for $r_0/L < 0.39858(3)$, the uniform phase becomes unstable to the so-called Gregory-Laflamme instability \cite{Gregory:1993vy}. Furthermore, from the onset of the instability, a new family of \emph{non-uniform} strings (middle panel of Fig.~\ref{fig:sketch}) emerges \cite{Horowitz:2001cz,Gubser:2001ac,Wiseman:2002zc,Sorkin:2006wp,Figueras:2012xj,Kalisch:2015via,Kalisch:2016fkm}. This phase still has an horizon whose spatial cross section has topology $S^1\times S^8$, but no longer exhibits translations invariance along the $Z$ direction.  We have constructed this phase using the DeTurck trick, confirming, and extending, the results reported in \cite{Figueras:2012xj}.

The last phase of interest to this letter (left-most panel in Fig.~\ref{fig:sketch}) corresponds to a solution where spatial cross sections of the event horizon no longer wrap the periodic direction, but instead have $S^9$ topology — these are often referred to as {\it localized} or {\it caged black holes}. Although this phase has been constructed in $d=5,6$ and $d=10$ (see \cite{Wiseman:2002ti,Sorkin:2003ka,Kudoh:2003ki,Kudoh:2004hs,Headrick:2009pv,Kalisch:2017bin,Dias:2017uyv,Ammon:2018sin,Cardona:2018shd}; see also \cite{Dias:2015pda,Dias:2016eto}), this has not been the case in $d=11$ (the dimension of interest in this letter). This phase is entirely novel and the details of its construction are given in the Supplemental Material.

All of the static phases satisfy a first law of black hole mechanics as well as a Smarr type relation, namely:
\begin{equation}
\mathrm d\mathcal{E}^{(0)}= \mathcal{T}_H^{(0)}\,\mathrm{d} \mathcal{S}_H^{(0)}\quad \text{and}\quad 8\,\mathcal{E}^{(0)} =9\, \mathcal{T}_H^{(0)}\,\mathcal{S}_H^{(0)}+\mathcal{T}_Z^{(0)} \,,
\label{SGStaticFirstlawSmarr}
\end{equation}
where $\mathcal{E}^{(0)}/G_{11}$, $\mathcal{S}_H^{(0)}/G_{11}$, $\mathcal{T}_Z^{(0)}/G_{11}$ are the energy, entropy and tension densities, respectively, while $\mathcal{T}_H^{(0)}/L$ is the Hawking temperature.

\paragraph*{\bf BFSS thermodynamics.}
Having provided some details on the solutions that can exist with standard Kaluza-Klein asymptotics, we now describe how to map these to the D0 brane asymptotics of~\eqref{eq:asym}. This is achieved by applying a Carrollian transformation of the form \footnote{This transformation can also be interpreted as a light-like compactification of the form
 $(T,Z) \to (T,Z) + (-R,+R)$, followed by a coordinate transformation and a particular choice of $R$. We are grateful to Juan~Maldacena for bringing this terminology to our attention.}
\begin{subequations}
\label{SG:shiftTransf}
\begin{equation}
T=\frac{L^{7/2}}{\mathcal{R}^{7/2}}\,\gamma\,t -  \frac{1}{\gamma}\,\frac{\mathcal{R}^{7/2}}{L^{7/2}}\,z\quad\text{and} \quad Z= \frac{1}{\gamma}\, \frac{\mathcal{R}^{7/2}}{L^{7/2}}\,\frac{z}{L}\,,
\end{equation}
where $\gamma$ is a constant to be adjusted in what follows so that the resulting metric has the correct asymptotic structure given in~\eqref{eq:asym}. It is a simple exercise to show that this transformation preserves regularity at the horizon, while modiying the asymptotic structure. This is crucial for the gravitational thermodynamics to directly correspond one-to-one with the thermodynamics of the BFSS dual theory. Using the general decaying properties of the several $Q_i$'s near spatial infinity, one finds
\begin{equation}
\gamma=\frac{1}{\pi ^{3/2}}\sqrt{\frac{15}{2}} \sqrt{\mathcal{E}^{(0)}-\mathcal{T}_Z^{(0)}}\,.
\end{equation}
\end{subequations}
As an example, the Carrollian transformation~\eqref{SG:shiftTransf}, takes the uniform black string~\eqref{eq:uni} to~\eqref{eq:larger}, which we recall describes a finite temperature stack of D0 branes.
Furthermore, since the static Kaluza-Klein circle has size $L$ at spatial infinity, and we require the M-theory circle $-$ parametrised by $z$ $-$ to have size $L_{11}=2\pi g_s \ell_s$ there, the transformation \eqref{SG:shiftTransf} uniquely determines $L$ to be:
\begin{equation}
L=N^{\frac{1}{9}}\left(32 \pi ^8\right)^{1/9}\,g_s^{1/3}\ell_s
\left(\mathcal{E}^{(0)} -\mathcal{T}_Z^{(0)}\right)^{-1/9}.
\end{equation}

\begin{figure}[tb]
    \centering
\includegraphics[width=0.36\textwidth]{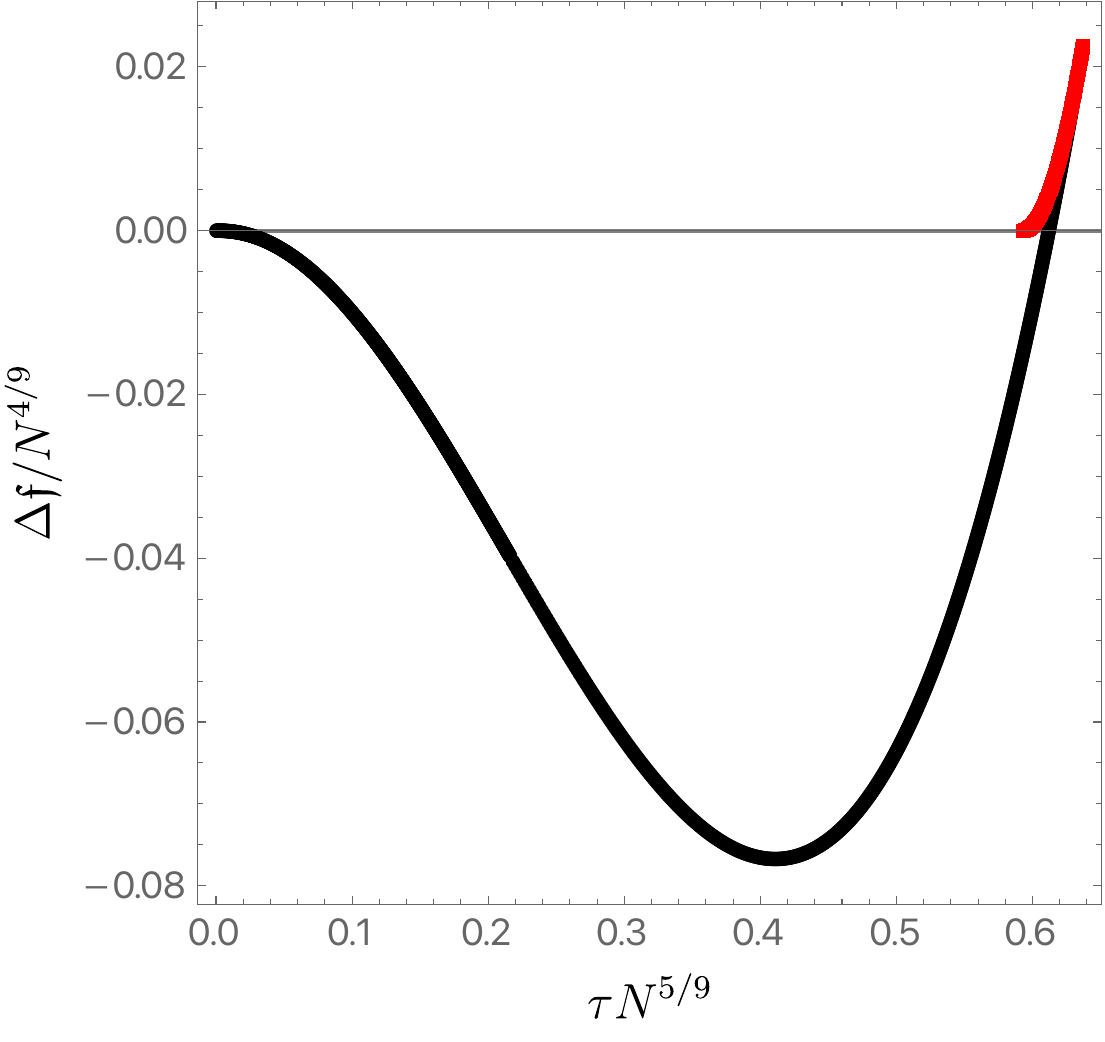}
    \caption{$\Delta \mathfrak{f}$, the free energy difference between a given thermal phase and the uniform phase at the same $\tau$, as a function of $\tau$. The black disks represent exact numerical data for the localized BFSS phase, while the red squares denote the non-uniform BFSS phase. 
    }
    \label{fig:canonical}
\end{figure}

Using \cite{Wald:1999wa,Dias:2019wof}, one can read off the energy $\varepsilon$, entropy $\sigma$, and free energy $\mathfrak{f}$ densities, as well as the temperature $\tau$ of all the BFSS field theory phases through the map (\ref{SG:shiftTransf}):
\begin{equation}\label{eq:QFTthermoMap}
\begin{split}
\varepsilon &=  N^{\frac{4}{9}} \left(32 \pi ^8\right)^{1/9}\,\frac{ \mathcal{E}^{(0)}+ \mathcal{T}_Z^{(0)} }{\left(\mathcal{E}^{(0)} -\mathcal{T}_Z^{(0)}\right)^{7/9}}, \\
\sigma &=  N  \, \frac{2\pi\, \mathcal{S}_H^{(0)}}{\mathcal{E}^{(0)}-\mathcal{T}_Z^{(0)}} \,,\\
\mathfrak{f} &= N^{\frac{4}{9}} \left(32 \pi ^8\right)^{1/9}
\,\frac{ \mathcal{E}^{(0)}+ \mathcal{T}_Z^{(0)}- 2 \,\mathcal{T}_H^{(0)} \,\mathcal{S}_H^{(0)}  }{\left(\mathcal{E}^{(0)} -\mathcal{T}_Z^{(0)}\right)^{7/9}}  \,,\\
\tau&=  N^{-\frac{5}{9}} \, \left(\frac{32}{\pi}\right)^{1/9}\, \mathcal{T}_H^{(0)} \left(\mathcal{E}^{(0)}-\mathcal{T}_Z^{(0)}\right)^{2/9} 
\end{split}
\end{equation}
respectively. These expressions, which will be derived in~\cite{Dias:2025jxx}, are central to the discussion in this letter. In particular, given a static (non-)uniform or localized gravitational solution, one can readily compute its thermodynamic properties $\{ \mathcal{E}^{(0)}, \mathcal{T}_Z^{(0)},\mathcal{T}_H^{(0)},\mathcal{S}_H^{(0)}\}$ and, via~\eqref{eq:QFTthermoMap}, obtain the corresponding field theory thermodynamics $\{\varepsilon,\sigma,\mathfrak{f},\tau\}$ of the (non-)uniform or localized BFSS phases.
\paragraph*{\bf Results.} 
We begin with the canonical ensemble, fixing the temperature $\tau N^{5/9}$ and identifying the phase with the lower free energy density, $\mathfrak{f}N^{-4/9}$. Fig.~\ref{fig:canonical} shows $\Delta \mathfrak{f}$, the free energy difference between a given BFSS thermal phase and the uniform phase at the same $\tau$. The black disks show exact numerical data for the localized BFSS phase\footnote{In a forthcoming publication~\cite{Dias:2025jxx}, we compare our numerical localized solutions with the second-order perturbative results from \cite{Harmark:2003yz} and find excellent agreement in regions of moduli space where localized black holes do exist.}
, and the red squares correspond to the non-uniform BFSS phase. We observe a first order phase transition between the localised phase and the uniform phase at $\tau N^{5/9}\approx 0.6118$. This is one of the main results of this letter. This aligns with the phase diagram conjectured by \cite{Itzhaki:1998dd}, but here we provide a precise prediction for the critical temperature where the first-order phase transition occurs. In principle, it should be possible to derive the latter from a field theory calculation. We have also computed the specific heat associated with all the dominant phases in the canonical ensemble and found it to be always positive. At first glance, this might seem at odds with expectations for small localized black holes. However, the map given in \eqref{eq:QFTthermoMap} ensures that the specific heat remains positive even for the localized black hole phase.

\begin{figure}[tb]
    \centering
    \includegraphics[width=0.36\textwidth]{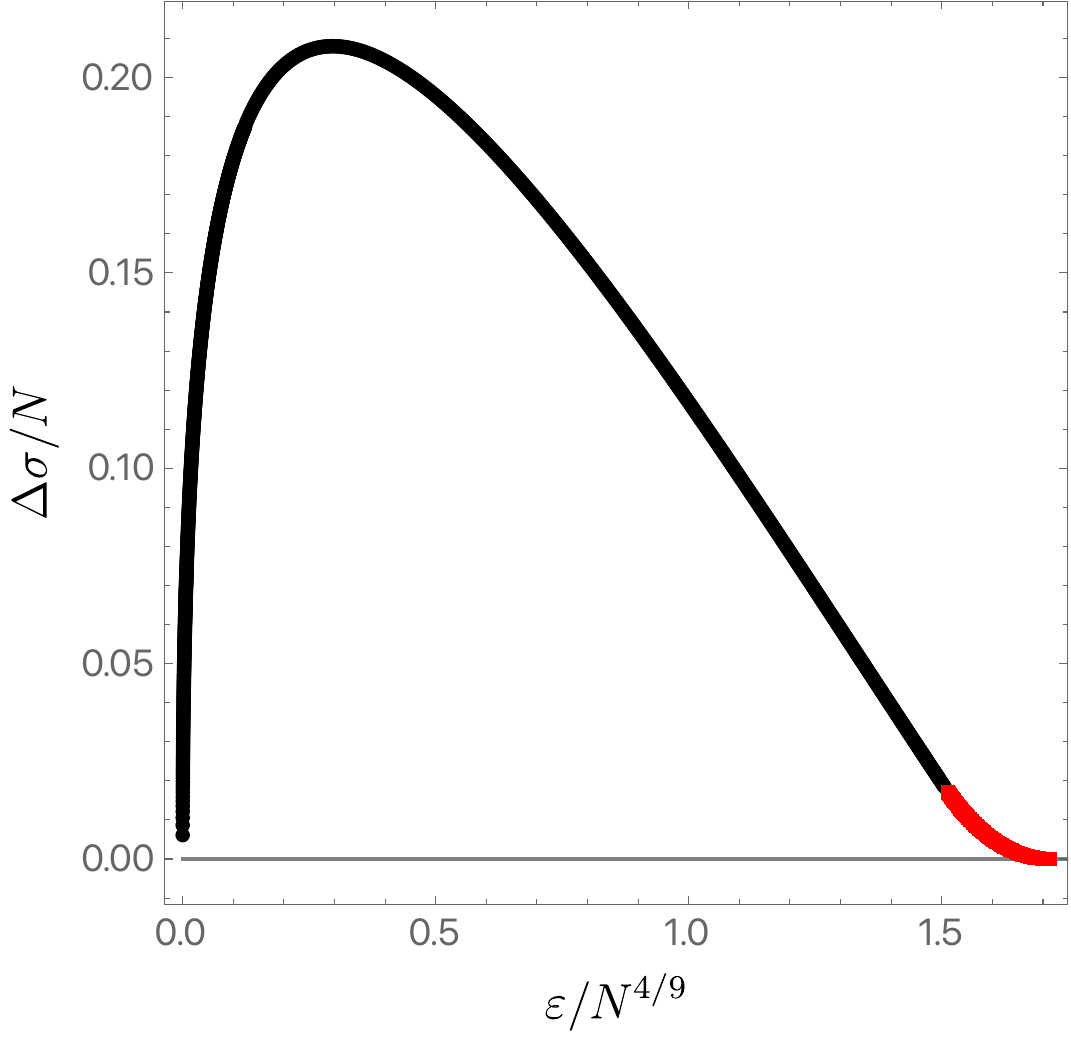}
    \caption{$\Delta \sigma$, the entropy density difference between a given thermodynamic phase and the uniform phase at the same energy $\varepsilon$, as a function of $\varepsilon$. The black disks represent exact numerical data for the localized BFSS phase, while the red squares denote the non-uniform BFSS phase.}
    \label{fig:micro}
\end{figure}

 On the other hand, the microcanonical ensemble reveals a much more interesting and unexpected structure. In Fig.~\ref{fig:micro}, using the same color coding as Fig.~\ref{fig:canonical}, we plot $\Delta \sigma/N$, the entropy density difference between a given BFSS phase and the uniform phase at fixed energy density $\varepsilon N^{-4/9}$. For  $0<\varepsilon< \varepsilon_{\rm m}$, the localized (black) phase dominates, while for $\varepsilon_{\rm m} <\varepsilon< \varepsilon_{\rm M}$, the non-uniform (red) phase dominates the microcanonical ensemble. A topology-changing transition occurs at $\varepsilon = \varepsilon_{\rm m}$, as described by Kol in \cite{Kol:2002xz}. One finds that $1.505(4)\lesssim\varepsilon_{\rm m}  N^{-4/9} \lesssim 1.512(6)$, where the lower bound is found computing the maximum value of $\varepsilon$ for which localized solutions exist, and the upper bound follows from finding the minimum value of $\varepsilon$ for which non-uniform solutions exist. Improved numerics would pin down $\varepsilon_{\rm m}$ even further. On the other hand, we find the value of $\varepsilon_{\rm M}$ to be $\varepsilon_{\rm M}  N^{-4/9} \approx 1.713(2)$. This simply follows from applying the map~\eqref{eq:QFTthermoMap} to the gravitational Gregory-Laflamme instability onset of the uniform phase.
 It remains a formidable challenge to predict $\varepsilon_{\rm m}$ and $\varepsilon_{\rm M}$ from a field theory calculation in the microcanonical ensemble and we believe our precise identification of these values should motivate and guide such a search. The existence of an energy window where the non-uniform phase dominates the microcanonical ensemble is surprising and constitutes the main result of this letter.
\paragraph*{\bf Discussion.} 
We have studied the low energy limit of BFSS using its holographic dual. Our numerical results obtained in the canonical ensemble corroborate, and make precise,  a longstanding conjecture set in \cite{Itzhaki:1998dd}. Namely, the dominant phase at small enough temperature is given by a localized thermal phase, dual to a black hole whose spatial horizon cross section has $S^9$ topology. The transition between the uniform BFSS phase and the localized BFSS phase occurs at $\tau N^{5/9}\approx 0.6118$. However, in the microcanonical ensemble we find a surprising result. For small enough energies, we find that the localized BFSS phase dominates, but we also find an energy \emph{window} where a non-uniform phase with $S^8\times S^1$ topology dominates, while breaking translational invariance along the M-theory circle.

One might wonder how our results can be made consistent with the dynamics of the original Gregory-Laflamme instability \cite{Gregory:1993vy,Lehner:2010pn,Figueras:2022zkg}, which show an instability of the uniform black string at sufficiently low energies. To address this question it is however important to note that Carrollian transformations~\eqref{SG:shiftTransf}  do not preserve dynamics, despite mapping stationary solutions. Specifically, the linear analysis in \cite{Gregory:1993vy} expands all metric functions in Fourier modes of the form $e^{-{\rm i}\,\omega\,T+2\pi n \, {\rm i}\,Z}$. It should be noted that~\eqref{SG:shiftTransf} maps these Fourier modes into modes in $t$ and $z$ that do \emph{not} respect the required periodicity in $z$. Recall that for the original Gregory-Laflamme unstable modes,  $\omega = {\rm i}\Omega$ with $\Omega>0$.

While there has been extensive research in high-energy physics aimed at understanding the canonical ensemble of various field theories $-$ using approaches such as localization, see for instance \cite{Bobev:2023ggk} and references therein $-$ it is only recently that the microcanonical ensemble has garnered significant attention \cite{Dias:2022eyq,Marolf:2022jra,Kim:2023sig}. The increased complexity of studying the microcanonical ensemble stems from the necessity of understanding the detailed nature of field theory microstates. Addressing this challenge is notoriously difficult and is a key reason why a complete solution to the black hole information paradox remains elusive. Indeed, recent calculations show how to compute the entropy of Hawking radiation (see for instance \cite{Penington:2019npb,Almheiri:2019psf,Almheiri:2019yqk,Almheiri:2019hni,Almheiri:2019psy}), but not how to compute its precise quantum state \cite{Almheiri:2020cfm}. Our work reveals an intricate microcanonical ensemble structure for BFSS quantum mechanics at low energies. It would be desirable to identify all the relevant phases using BFSS degrees of freedom directly, though this task will invariantly require an understanding of the underlying microstates.

One possible interpretation for the absence of a first-order phase transition in the microcanonical ensemble, typically observed in phase transitions between localized black holes and uniform phases in $d \leq 13$ \cite{Wiseman:2002ti,Sorkin:2003ka,Kudoh:2003ki,Kudoh:2004hs,Headrick:2009pv,Kalisch:2017bin,Dias:2017uyv,Ammon:2018sin,Cardona:2018shd} with standard Kaluza-Klein asymptotics, is as follows. In standard thermodynamic systems, first-order phase transitions in the microcanonical ensemble can lead to mixed (also known as coexistent) phases. However, the boundary system in this context is quantum mechanical, which prevents the spatial separation of phases. Instead, we observe the presence of an intermediate phase (the non-uniform black string) that effectively interpolates between the localized and uniform phases.\footnote{We would like to thank Juan~Maldacena for suggesting this interpretation.}

The Carrollian map~\eqref{SG:shiftTransf} together with the associated thermodynamical map \eqref{eq:QFTthermoMap} allows us to associate \emph{any} black hole solution with standard Kaluza-Klein asymptotics into one with D0 brane asymptotics (and thus with any BFSS state). In particular, given the multitude of black hole solutions in higher dimensions $-$ see, for instance, \cite{Emparan:2008eg,Dias:2010eu,Dias:2022mde,Dias:2022str,Dias:2023nbj} $-$  and the nontrivial nature of the map (\ref{eq:QFTthermoMap}), one wonders how many more unexpected phases can possibly dominate a generalized ensemble that includes rotations and/or supersymmetric mass deformations \emph{\`a la} BMN \cite{Berenstein:2002jq,Costa:2014wya}. We leave these exploratory investigations to future work.

\paragraph*{Acknowledgements.}
We would like to thank Gary~Horowitz, Juan~Maldacena and Yoav~Zigdon for reading an earlier version of this letter and for providing critical comments.  O.D. acknowledges financial support from the STFC ``Particle Physics Grants Panel (PPGP) 2018" Grant No.~ST/T000775/1 and PPGP 2020 grant No.~ST/X000583/1. O.D.'s research was also supported in part by the YITP-ExU long-term workshop {\it Quantum Information, Quantum Matter and Quantum Gravity (QIMG2023)}, Yukawa Institute for Theoretical Physics, Kyoto Univ., where part of this work was completed. J. E. S. has been partially supported by STFC consolidated grants ST/T000694/1 and ST/X000664/1.

\newpage
\bibliography{refsLocalizedBFSS}


\clearpage
\onecolumngrid
\appendix

\setcounter{equation}{0}
\renewcommand\theequation{A.\arabic{equation}}

\section{Constructing the several black hole phases}
In this appendix, we will stay within the language of 11-dimensional vacuum Einstein gravity or, equivalently, of 11-dimensional supergravity with vanishing 3-form gauge potential, $C=0$. We numerically construct the localised black hole and non-uniform string phases, i.e. the vacuum Einstein solutions sketched in the left and middle panels of Fig.~\ref{fig:sketch}.   As for the uniform black string (right panel of Fig.~\ref{fig:sketch}) these solutions have  ${\mathbb R}^{(1,9)}\times S_L^1$ asymptotics where the $S_L^1$ has circumference $L$. The non-uniform strings contain a horizon that covers the entire $S_L^1$, while the $S_L^1$ is partially exposed in localised black hole case.

We use the Einstein-de Turck formulation of the gravitational equations of motion \cite{Headrick:2009pv,Figueras:2011va,Wiseman:2011by,Dias:2015nua}. This formulation requires that we first choose a reference metric $\overline g$.  This metric need not be a solution to the Einstein equation, but must contain the same symmetries and causal structure as the desired solution and, to facilitate the numerical search,  should resemble the final solution near the horizon and asymptotically.  With the reference metric chosen, the DeTurck method then modifies the Einstein equation $R_{\mu\nu}=0$ to \eqref{eq:ede}-\eqref{eq:ede2}. Unlike $R_{\mu\nu}=0$, the Einstein-de Turck equation yields PDEs that are elliptic in character and solutions of the Einstein-de Turck system are also solutions of  11-dimensional vacuum Einstein gravity as long as the de Turck vector vanishes, $\xi_\mu\xi^\mu=0$, which is what we want. 

\subsection{Localized black holes}

In this subsection we want to construct the localised black holes of 11-dimensional vacuum Einstein gravity.   
A perturbative construction of these solutions is available  \cite{Harmark:2003yz,Gorbonos:2004uc,Dias:2007hg} but only for small energies. Since we also need to know the properties of localized black holes at high energies, we will employ numerical methods to solve the equations of motion. In~\cite{Dias:2025jxx}, we shall compare our numerical localized solutions with the second-order perturbative results from \cite{Harmark:2003yz} and find excellent agreement in regions of moduli space where localized black holes do exist.

Localized black holes are static, axisymmetric, and asymptotically ${\mathbb R}^{(1,9)}\times S^1_L$ black holes with horizon topology $S^9$. Within the Einstein-de Turck formution of the equations of motion, a de Turck reference metric tailored to find such solutions must therefore contain an axis, a topologically $S^9$ horizon, and asymptote to ${\mathbb R}^{(1,9)}\times S^1_L$. Moreover, the periodic  $S^1_L$ coordinate has a $\mathbb Z_2$ symmetry, which we use in our benefit to halve the integration domain along this direction. For very small energies, the solution near the horizon is expected to resemble the 11-dimensional Schwarzschild-Tangherlini with a round $S^9$. But was the energy increases one expects that the $S^9$ gets increasingly deformed.
 To accommodate the five boundaries of the system, it is convenient to work with  two different patches, each one with four boundaries and their own coordinate system. One of these patches and associated coordinate chart is adapted to the asymptotic region, and the other to the near horizon region.

To engineer our reference metric in the coordinate chart adapted to the asymptotic region, we begin by analysing the system near infinity. We start with the ${\mathbb R}^{(1,9)}\times S^1_L$ solution
\begin{equation}\label{dsflat}
\mathrm ds^2_{{\mathbb R}^{(1,9)}\times S^1_L}=-{\mathrm d}\mathcal{T}^2+\mathrm dr^2+r^2{\mathrm d}\Omega_8^2+L^2 dZ^2\;,
\end{equation}
where $Z \in(-\frac{1}{2},\frac{1}{2})$ is the periodic coordinate that parametrizes the circle $S^1_L$ with length $L$. Next, to scale out $L$ and work with compactified coordinates, we  use the change of coordinates  $T=\frac{\pi}{L}\, \mathcal{T}$, $r=\frac{\pi}{L}\rho\sqrt{2-\rho^2}/(1-\rho^2)$  and $Z=2 \pi \arcsin(\xi/\sqrt 2)$,   so that \eqref{dsflat} now reads
\begin{equation}
\mathrm ds^2_{{\mathbb R}^{(1,9)}\times S^1_L}=\frac{L^2}{\pi^2}\bigg[-{\mathrm d}\mathcal{T}^2+\frac{4\mathrm d\rho^2}{(2-\rho^2)(1-\rho^2)^4}+\frac{4\mathrm d\xi^2}{2-\xi^2}+\frac{\rho^2(2-\rho^2)}{(1-\rho^2)^2}{\mathrm d}\Omega_8^2\bigg]\;.
\end{equation}
The coordinate ranges are now $\rho\in[0,1]$ and $\xi\in[-1,1]$.  Exploiting the $\mathbb Z_2$ symmetry in $\xi$, onwards we take $\xi\in[0,1]$ and require reflection symmetry at $\xi=0$ and $\xi=1$. The origin is at $\rho=0$ and the asymptotic infinity is at $\rho=1$.  

This analysis invites us to choose the following reference metric in the far region:
\begin{equation}\label{reffar}
\overline{\mathrm ds}^2=\frac{L^2}{\pi^2}\bigg[-m\,{\mathrm d}\mathcal{T}^2+g\bigg(\frac{4\mathrm d\rho^2}{(2-\rho^2)(1-\rho^2)^4}+\frac{4\mathrm d\xi^2}{2-\xi^2}+\frac{\rho^2(2-\rho^2)}{(1-\rho^2)^2}{\mathrm d}\Omega_8^2\bigg)\bigg]\;,
\end{equation}
where $m$ and $g$ are functions of $\rho$ and $\xi$ that we must and will specify later. 
The motivation to choose this reference metric is three-folded. First, the reference metric must have the same causal structure and symmetries as the localized black hole solution we look for. This is the case if $m(\rho,\xi)$ is chosen to have a horizon (more later). Second, the $\mathrm d\rho^2$ and $\mathrm d\xi^2$ components in the metric originate directly from $\mathrm dr^2+L^2 \mathrm dZ^2$ that is manifestly flat. Thus, using conformal mappings, we can transform  \eqref{reffar} straightforwardly into other orthogonal coordinates. This will allow us to select a coordinate transformation whereby, together with a natural choice of functions $m$ and $g$,  the reference geometry  describes a black hole. Finally, when the black hole has small energy (high temperature), we would like the reference metric close to the horizon to resemble asymptotically flat 11-dimensional Schwarzschild-Tangherlini, as expected of the localized solution we search for. The reference metric in the form \eqref{reffar} is ready to accommodate this requirement if we  write Schwarzschild-Tangherlini in isotropic coordinates (as we do below).

\begin{figure}[ht]
\centering
\includegraphics[width=.5\textwidth]{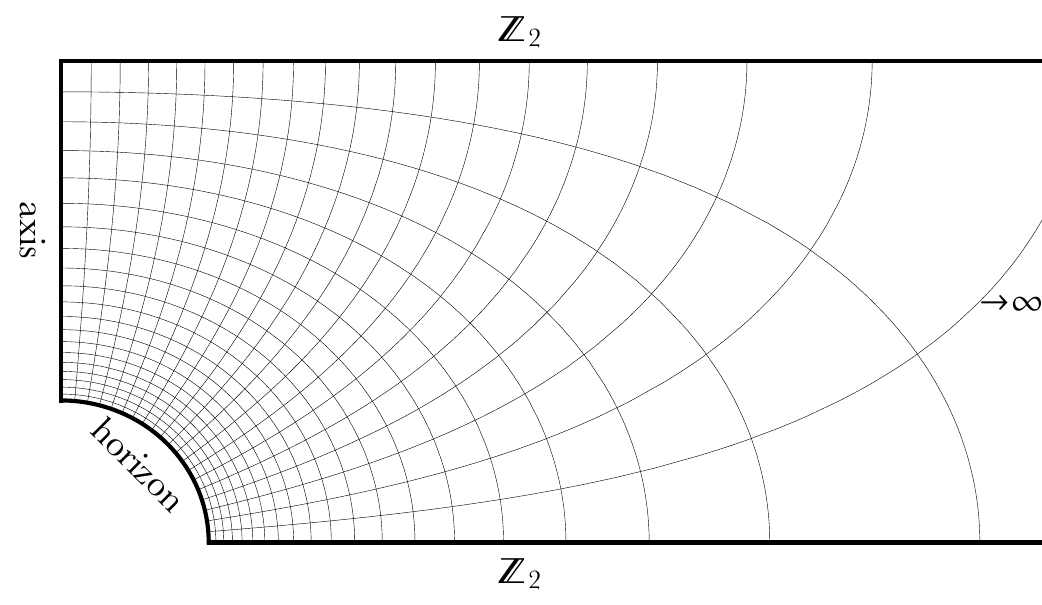}
\caption{Sketch of integration domain in $\{r,Z\}$ coordinates. The $\{\rho,\xi\}$ coordinates we use in the far region are related to these coordinates through $r=\frac{\pi}{L}\rho\sqrt{2-\rho^2}/(1-\rho^2)$  and $Z=2 \pi \arcsin(\xi/\sqrt 2)$.  The grid lines are lines of constant $x$ and constant $y$ where $\{x,y\}$ are the coordinates used near the horizon.}\label{Fig:coord}
\end{figure}

The far-region coordinates $(\rho,\xi)$ are effectively elliptic coordinates that are sometimes used to describe 2-dimensional flat space. This 2-dimensional space can also be written in bipolar coordinates  $(x,y)$. 
Both of these coordinate systems can be obtained from a conformal mapping of 2-dimensional Cartesian coordinates. These conformal mappings also allow to obtain the coordinate transformation, and its inverse,  between the eliptic and bipolar coordinates:
\begin{eqnarray}\label{coordmap}
   && x=\sqrt{1-\frac{\sinh {\bigl( }\frac{\rho  \sqrt{2-\rho ^2}}{1-\rho ^2}{\bigr) }}{\sqrt{\xi^2\left(2-\xi ^2\right)+\sinh^2{\bigl( }\frac{\rho  \sqrt{2-\rho^2}}{1-\rho ^2}{\bigr) }}}},  \qquad y=\frac{y_0\left(1-\xi ^2\right)}{\sqrt{\xi^2\left(2-\xi ^2\right)+\sinh^2{\bigl( }\frac{\rho  \sqrt{2-\rho^2}}{1-\rho ^2}{\bigr) }}} \,;\nonumber \\
   && \rho=\sqrt{1-\frac{1}{\sqrt{1+{\rm arcsinh}^2{\bigl( }\frac{y_0 \left(1-x^2\right)}{\sqrt{y^2+y_0^2  \, x^2 \left(2-x^2\right)}}{\bigr) }}}}, \qquad \xi=\sqrt{1-\frac{y}{\sqrt{y^2+y_0^2 \, x^2 \left(2-x^2\right)}}}\,.
\end{eqnarray}
 In these new coordinates  $(x,y)$, the de Turck reference metric \eqref{reffar} becomes
\begin{align}\label{refnear}
      \overline{ds}^2= \frac{L^2}{\pi^2} {\bigg \{} -m\,{\mathrm d}\mathcal{T}^2+g {\biggl [} \frac{y_0^2}{h} \left( \frac{{\mathrm d}y^2}{y^2+y_0^2} +\frac{4\mathrm dx^2}{2-x^2}\right) +\, s \left(1-x^2\right)^2  {\mathrm d}\Omega_8^2 {\biggr ]}  {\bigg \}}\;,
\end{align}
where we assume $y_0>0$, $m$ and $g$ transform as scalars,  and we defined
\begin{equation}\label{hsdef}
 h=y^2 + y_0^2 x^2 \left(2-x^2\right) \,,\qquad s=\frac{{\rm arcsinh}^2{\bigl (}\frac{y_0 \left(1-x^2\right)}{\sqrt{y^2+y_0^2 \,x^2 \left(2-x^2\right)}}{\bigr )}}{\left(1-x^2\right)^2}.
\end{equation}
The function $s$ is positive definite and regular, even at $x=1$, while $h$ is positive except at $x=y=0$, where it vanishes. The axis is at $x=1$ and asymptotic infinity is at the coordinate `point' $x=y=0$.  The locations $\xi=0$ and $\xi=1$, where we require reflection symmetry, are mapped to $x=0$ and $y=0$, respectively. The location $\rho=0$, $\xi=0$ is mapped to $y\to\infty$. Anticipating the fact that we want to place a horizon at $y=1$, we have introduced the constant $y_0$ in \eqref{coordmap} and thus in \eqref{refnear}-\eqref{hsdef}.  That is, $y_0$ moves the location of this horizon in the $\{\rho, \xi\}$ coordinates and its value effectively determines the black hole size (and temperature). The integration domain and grid lines of constant $x$ and $y$ are sketched in Fig. \ref{Fig:coord}.

We  now describe the motivation that leads to our choice of the functions $m$ and $g$ that appear in \eqref{reffar} and \eqref{refnear}.  We have some freedom to choose these functions but they must satisfy a few requirements. To have the required asymptotics, they must approach $1$ at $x=y=0$.  The reflection symmetries of the problem requires that $m$ and $g$ are even functions of $y$ and $x$. Finally, one must have a regular horizon at $y=1$. Moreover, ideally, we would like the geometry near the horizon for small $y_0$ to resemble asymptotically flat Schwarzschild-Tangherlini in isotropic coordinates, which in $d$-dimensions can be written as
\begin{equation}\label{schwiso}
\mathrm ds^2_{\mathrm{Schw}}=-\left(\frac{1-y^{d-3}}{1+y^{d-3}}\right)^2{\mathrm d}\mathcal{T}^2+y_0^2(1+y^{d-3})^{\frac{4}{d-3}}\left(\frac{\mathrm dy^2}{y^4}+\frac{1}{y^2}\mathrm d\Omega_{d-2}^2\right)\;.
\end{equation}
Note that \eqref{schwiso} look similar to the $y_0\to0$ limit of our reference metric \eqref{refnear}, a fact that motivated the choice of the latter. This motivates the choice
\begin{equation}\label{mgdef}
m=\frac{1}{1+y_0^2x^2(2-x^2)}\left(\frac{1-y^6}{1+y^6}\right)^2\;,\qquad g=(1+y^6)^{2/3}\;.
\end{equation}
Note that we used $d=9$ components from \eqref{schwiso} rather than $d=11$ because it turns out to yield better numerics (while still obeying the requirement that $m$ and $g$ are even functions of $y$; note that the functions $\tilde{f}_i$ and $f_i$ to be defined below capture the 11-dimensional nature of the solution). The extra $x$ dependence in the numerator of $m$ is placed to fix regularity of the horizon. These functions $m,g$ can be mapped to $\{\rho,\xi\}$ coordinates straightforwardly via \eqref{coordmap}.  

Summarizing, we have chosen a de Turck reference metric in two two coordinate systems \eqref{reffar}, \eqref{refnear} whereby the auxiliary functions given in \eqref{hsdef} and \eqref{mgdef}, and the far and near region coordinates are related by the map \eqref{coordmap}. This reference metric provides the structural skeleton for the final metric ansatz:
\begin{align}\label{localisedansatz}
      {\mathrm ds}^2 &= \frac{L^2}{\pi^2} {\biggl \{} -m \tilde{f}_1{\mathrm d}\mathcal{T}^2+g {\biggl [} \frac{4 \tilde{f}_2\,{\mathrm d}\rho^2 }{(2-\rho^2)(1-\rho^2)^4}
+\frac{4\tilde{f}_3}{2-\xi^2}\left( d\xi -\tilde{f}_5 \frac{\xi(2-\xi^2)(1-\xi^2)\rho}{(1-\rho^2)^2}d\rho \right)^2
+\tilde{f}_4\frac{\rho^2(2-\rho^2)}{(1-\rho^2)^2} \, {\mathrm d}\Omega_8^2  {\biggr ]}    {\biggr \}},\nonumber\\
&= \frac{L^2}{\pi^2} {\Biggl (} -m\,f_1\,{\mathrm d}\mathcal{T}^2+g {\biggl \{} \frac{y_0^2}{h} \bigg[ \frac{f_2\,{\mathrm d}y^2}{y^2+y_0^2} +\frac{4 f_3}{2-x^2} \bigg( {\mathrm d}x -f_5\,\frac{x \left(2-x^2\right) \left(1-x^2\right) y \left(1-y^2\right)}{h}\,{\mathrm d}y  \bigg)^2\bigg] \nonumber \\
      & \quad\quad\quad\quad\quad\quad\quad\quad\quad\quad +f_4\, s \left(1-x^2\right)^2  {\mathrm d}\Omega_8^2 {\biggr \}}  {\Biggr )}\;,
\end{align}
The known functions $\{m,g,h,s\}$ are treated as scalars, transforming between the far and near  coordinate systems as \eqref{coordmap}. One the other hand, $\tilde f_i$ are unknown functions of $\{\rho,\xi\}$, and $f_i$ are unknown functions of $\{x,y\}$ that we need to solve for. When we set ${\tilde f}_{1,2,3,4}=1, {\tilde f}_5=0$ and  $f_{1,2,3,4}=1, f_5=0$ we recover the reference metric \eqref{reffar} and \eqref{refnear}, respectively. 

\begin{figure}[ht]
\centering
\includegraphics[width=.35\textwidth]{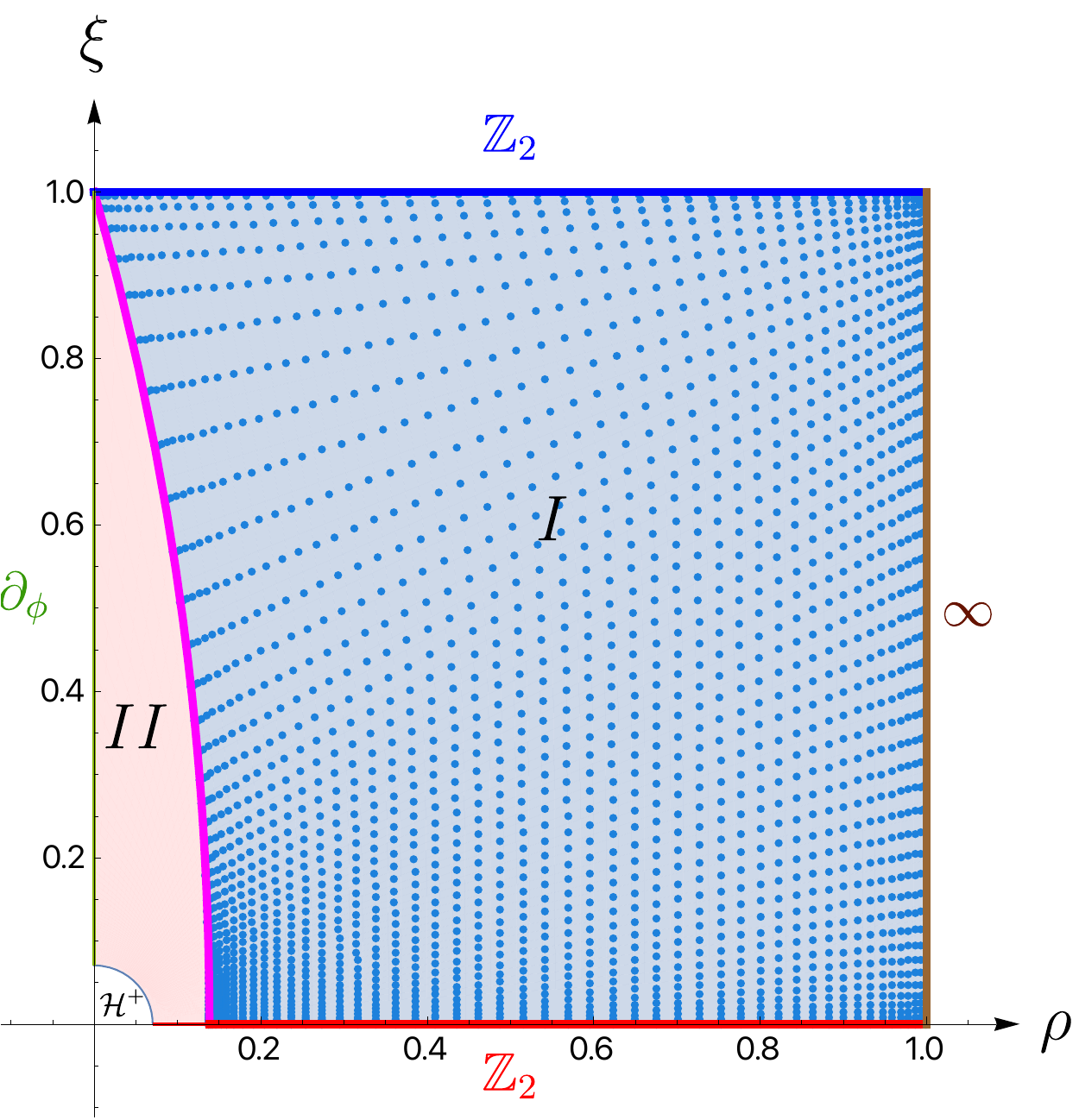}\hspace{1cm}
\includegraphics[width=.35\textwidth]{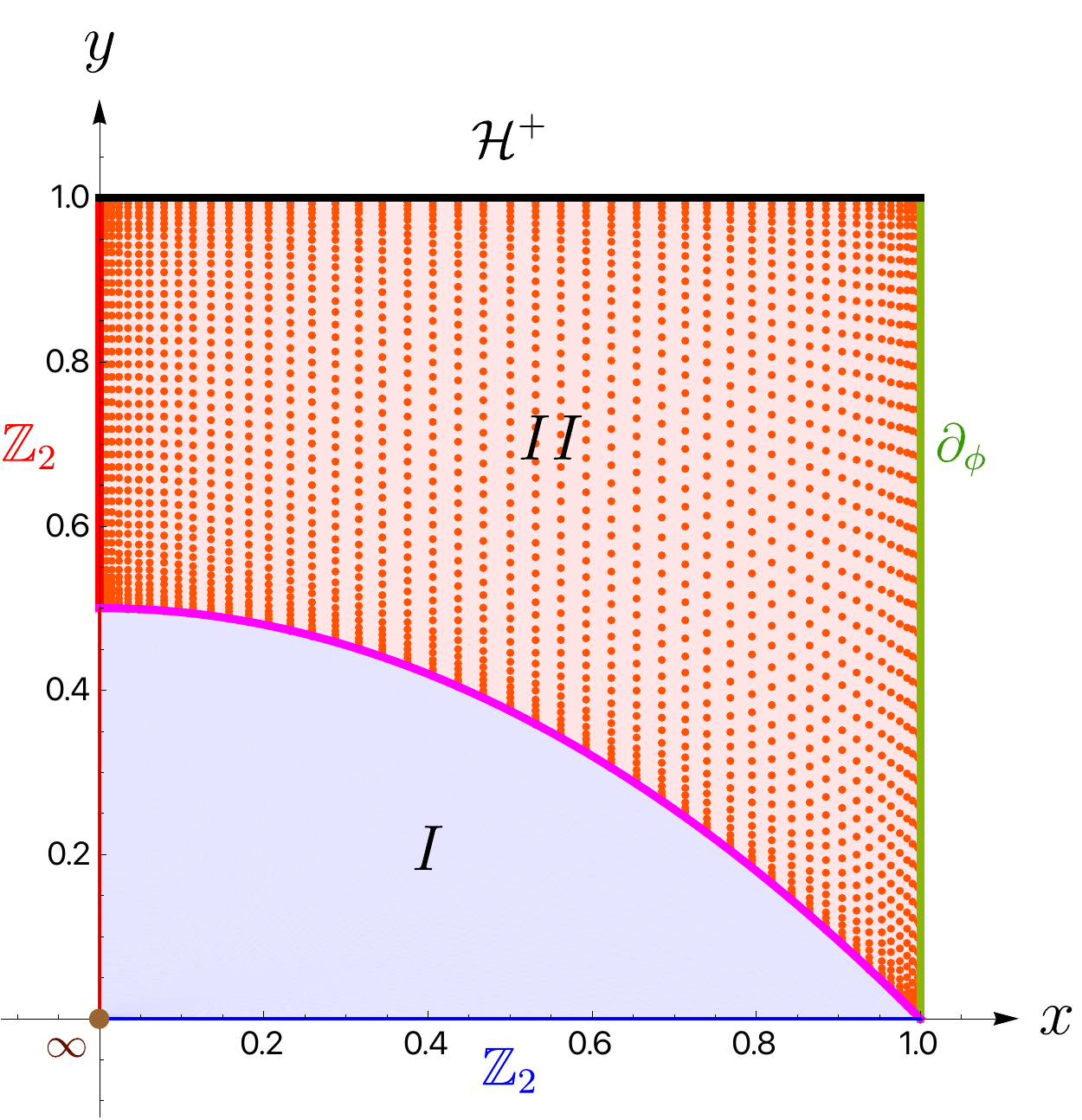}
\caption{Integration domain with two patches $I$ and $II$.  Chebyshev-Gauss-Lobatto grids with $50\times 50$ points are placed using transfinite interpolation. {\bf Left panel:} Patch $I$ (in the far region) uses  $\{\rho,\xi\}$ coordinates and patch $II$ (near the horizon in the quarter circle in the lower left) is mapped from $\{x,y\}$ coordinates using \eqref{coordmap}.  {\bf Right panel:} Patch II (near the horizon) uses  $\{x,y\}$ coordinates.}\label{Fig:patches}
\end{figure}

The Einstein-de Turck equations motions $-$ \eqref{eq:ede} in the main text $-$ for \eqref{localisedansatz} must be solved together with the boundary conditions that we now discuss. At the asymptotic boundary $\rho=1$ we impose Dirichlet conditions ${\tilde f}_{1,2,3,4}=1, {\tilde f}_5=0$: the solution must approach the reference metric.  At the horizon $y=1$, regularity of the solution  requires that $f_i$ obey certain Robin boundary conditions (whose expressions  are long and unilluminating). At the remaining boundaries, $f_i$ and $\tilde{f}_i$ must obey Neumann conditions either due to regularity at the axis $\rho=0$ (or $x=1$), or reflection symmetry at $\xi=0$ (or $x=0$) and at $\xi=1$ (or $y=0$).

We solve the boundary value problem numerically using a Newton-Raphson algorithm.  To discretise the numerical grid,  we divide the integration domain into two patches as shown in Fig.~\ref{Fig:patches}: in the far region patch $I$ we use  $\{\rho,\xi\}$ coordinates while in the near horizon patch $II$ we use  $\{x,y\}$ coordinates. In each patch we place Chebyshev-Gauss-Lobatto $N \times N$ grids using transfinite interpolation (these methods are reviewed in \cite{Dias:2015nua}; the results we present have $N=50$). The patching boundary between patches I and II (magenta curve) is given by $y=k_0 (1-x^2)$, where we have freedom to choose $k_0=1/2$ (the value we used to get our numerical results). 
At this patch boundary, we require that the line elements given by \eqref{localisedansatz} do match, and we also require that the normal derivative across the patch boundary do match. 

Given $f_i(x,y)$, the entropy and temperature of the localised black holes are: 
\begin{equation}\label{thermoLocGRnum}
\mathcal{S}^{(0)}_H=\frac{L^9}{G_{11}} \int_0^1 \mathrm dx\,\frac{256\,y_0 \,{\rm arcsinh}^8{\bigl( }\frac{y_0 \left(1-x^2\right)}{\sqrt{1+y_0^2\,x^2 \left(2-x^2\right)}} {\bigr) }}{105 \pi^5 \sqrt{2-x^2}\sqrt{1+y_0^2\,x^2 \left(2-x^2\right)}} \,\sqrt{f_3(x,1)}\, f_4(x,1)^4\,,\qquad\qquad 
\mathcal{T}^{(0)}_H=\frac{1}{L}\,\frac{3}{2^{4/3}}\frac{\sqrt{1+y_0^2}}{y_0}.
\end{equation}
Localized black holes are a one-parameter family of solutions that we can take to be parametrized by the temperature $\mathcal{T}^{(0)}_H$, that is to say, by $y_0$ for our numerical solutions.
 $L$ just sets a scale and it drops out of the equations of motion. 
As a first seed for the Newton-Raphson algorithm, we choose the reference metric with $y_0=1/10$ and then march our code for increasing or decreasing values of $y_0$. 
Our selection of reference metric has restricted our temperature range to $T\,L>3/2^{4/3}\approx 1.19$; this is \eqref{thermoLocGRnum} with $y_0=0$. Our solutions have temperature above this critical value. To get the Helmoltz free energy $\mathcal{F}^{(0)}$ we integrate the first law of thermodynamics, $\mathrm d\mathcal{F}^{(0)}=-\mathcal{S}^{(0)}_H \,\mathrm d\mathcal{T}^{(0)}_H$, and the energy is then $\mathcal{E}^{(0)}=\mathcal{F}^{(0)}+\mathcal{T}^{(0)}_H \mathcal{S}^{(0)}_H$.  The tension $\mathcal{T}_Z^{(0)}$ along the circle $S^1_L$ can be obtained from the Smarr relation \eqref{SGStaticFirstlawSmarr}, $ 8\,\mathcal{E}^{(0)} =9\, \mathcal{T}_H^{(0)}\,\mathcal{S}_H^{(0)}+\mathcal{T}_Z^{(0)}$.
The values of these thermodynamics quantities are in excellent agreement with the results obtained using the standadard Arnowitt-Deser-Misner formalism \cite{Harmark:2003yz,Dias:2007hg} or the covariant Noether charge formalism (a.k.a. covariant phase space method) \cite{Wald:1999wa,Dias:2019wof}. 

The BFSS thermodynamics $\{ \varepsilon, \sigma, \tau, \mathfrak{f} \}$ can then be obtained from $\{ \mathcal{E}^{(0)}, \mathcal{T}_Z^{(0)},\mathcal{T}_H^{(0)},\mathcal{S}_H^{(0)}\}$ via the map \eqref{eq:QFTthermoMap} in the main text.

\subsection{Non-uniform strings}
Constructing the non-uniform solutions is much simpler than constructing the localized black holes. In particular, the domain of integration naturally fits within a square, allowing us to use a single coordinate system to cover the entire integration domain. Our Ansatz for solving the Einstein-DeTurck equation reads
\begin{subequations}
\begin{equation}
{\rm d}s^2=L^2\left[-G(x)x^2 Q_1{\rm d}\tilde{T}^2+\frac{4\,y_+^2\,Q_2 {\rm d}x^2}{G(x)(1-x^2)^4}+\frac{y_+^2\,Q_5}{(1-x^2)^2}{\rm d}\Omega^2_8+Q_4\left({\rm d}y+Q_3 {\rm d}x\right)^2\right]
\end{equation}
where
\begin{equation}
G(x)=\sum_{i=0}^6(1-x^2)^i\,,
\end{equation}
\end{subequations}%
and the functions $Q_i$ are to be determined via the Einstein-DeTurck equation and depend only on $x$ and $y$. As a reference metric, we take
\begin{equation}
Q_1=Q_2=Q_4=Q_5=1\quad\text{and}\quad Q_3=0\,
\end{equation}
which we recognize as the uniform black string (\ref{eq:uni}) in the main text under the identification $X=x$, $Z=y$, $y_+\equiv r_0/L$ and $\tilde{T}=T/L$.

We now address the challenging issue of boundary conditions. At spatial infinity we demand that all the functions approach those representing the reference metric, while at $y=0$ and $y=1/2$ we demand that
\begin{equation}
\left.\frac{\partial Q_i}{\partial y}\right|_{y=0,1/2}=0\,,\quad\text{for}\quad i=1,2,4,5\,, \qquad\qquad Q_3\big|_{y=0,1/2}=0\,.
\end{equation}
Finally, regularity at the bolt $x=0$ demands that
\begin{equation}
\left.\frac{\partial Q_i}{\partial x}\right|_{x=0}=0\,,\quad\text{for}\quad i=1,2,4,5\,, \qquad\qquad Q_3\big|_{x=0}=0\,,
\end{equation}
which also imply $Q_1(0,y)=Q_2(0,y)$ via regularity.

Although the problem is relatively simple to setup as a well defined PDE system, it turns out to be rather difficult to solve numerically when the black strings are very deformed. This occurs because non-uniform strings become highly distorted near the Kol merger, as anticipated. Indeed, near the merger, the black string wants to pinch the $S^8$, transitioning to a localized black hole with $S^9$ topology. Since our domain of integration naturally lives on a square $(x,y)\in(0,1)\times(0,1/2)$, the pinch of the black string will occur either at $(x,y)=(0,0)$ or $(x,y)=(0,1/2)$, with both solutions being equivalent. For dominant non-uniform strings (note that excited strings can have several nodes) these are the only two possibilities. Near the pinch, the solutions develop very large gradients that have to be dealt with great care.

\begin{figure}[ht]
\centering
\includegraphics[width=.6\textwidth]{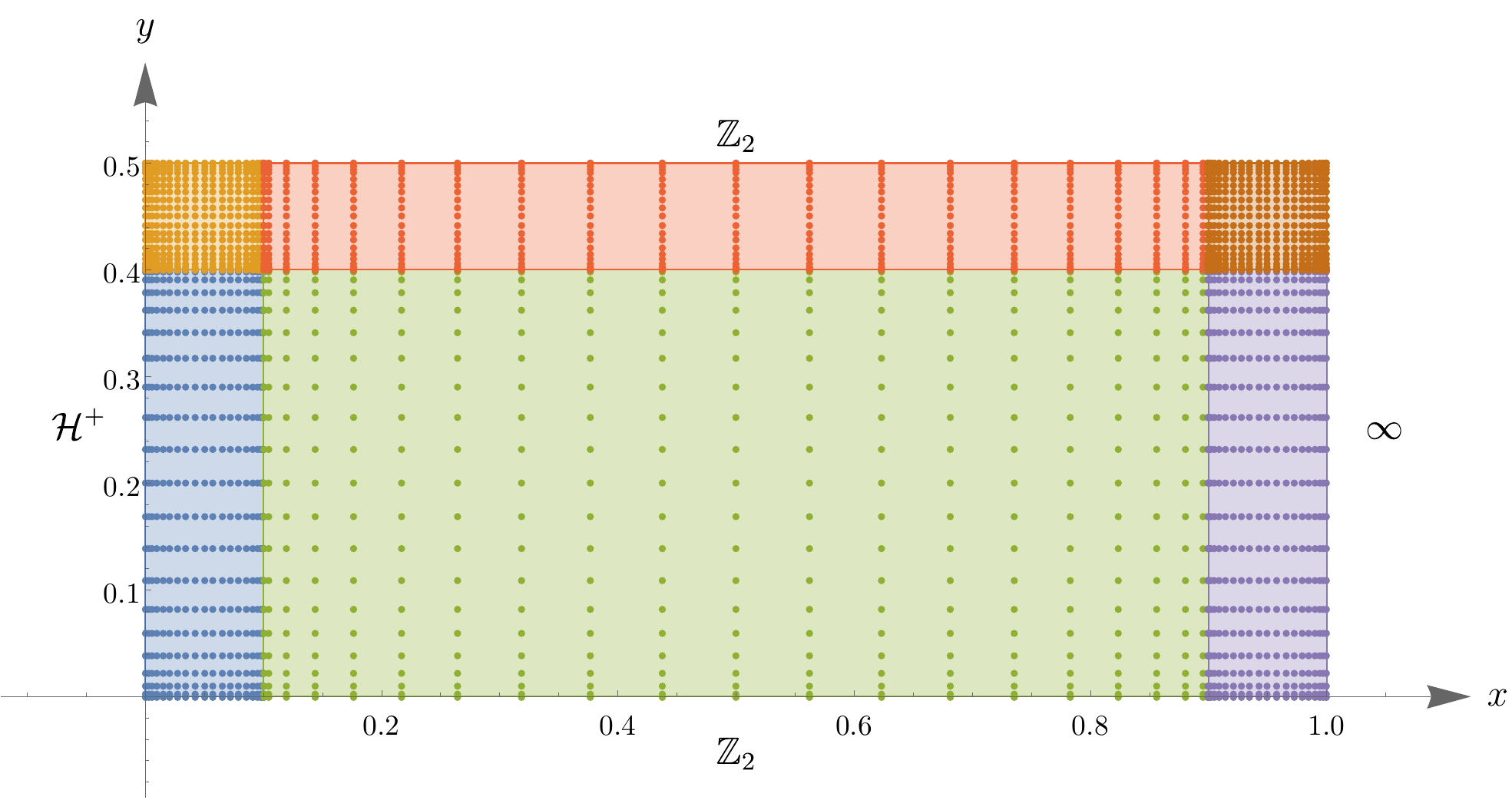}
\caption{Integration domain for the non-uniform strings, highlighting the various patches and regions where higher resolution is required.}\label{fig:appen}
\end{figure}

To deal with the large gradients, we break the domain of integration into six subdomains and use Coon maps, as detailed in \cite{Dias:2015nua}, to link the several pacthes together. In Fig.~\ref{fig:appen} we show the several patches and this figure assumes that the pinching is ocurring near $(x,y)=(0,1/2)$. This is the corner where one needs high resolution to resolve the large gradients.

Finally, reading off the energy near spatial infinity is notoriously difficult in higher dimensions. To address this, we use a different set of variables in the last two patches near $x=1$. To motivate this change, we first present the asymptotic expansion of all the $Q_i$ near spatial infinity by solving the Einstein DeTuck equation asymptotically. This expansion is as follows:
\begin{align}
Q_1(x,y)&=1+(1-x)^7 \delta _1-\frac{7}{2} (1-x)^8 \delta _1+\frac{21}{4} (1-x)^9 \delta _1+\ldots\nonumber
\\
Q_2(x,y)&=1+(1-x)^7 \delta _2-\frac{7}{2} (1-x)^8 \delta _2-8 (1-x)^9 \delta _5+\ldots\nonumber
\\
Q_3(x,y)&=(1-x)^6 \delta _3-3 (1-x)^7 \delta _3+\frac{15}{4} (1-x)^8 \delta _3-\frac{5}{2} (1-x)^9 \delta _3+\ldots
\\
Q_4(x,y)&=1+(1-x)^7 \delta _4-\frac{7}{2} (1-x)^8 \delta _4+\frac{21}{4} (1-x)^9 \delta _4+\ldots\nonumber
\\
Q_5(x,y)&=1+(1-x)^7 \delta _2-\frac{7}{2} (1-x)^8 \delta _2+\left(\delta _5+\frac{189 \delta _2}{32}\right) (1-x)^9+\ldots\nonumber
\end{align}
where the $\ldots$ has two types of contributions. First, we have higher order polynomial contributions of the form $(1-x)^{n}$ with $n\geq10$. Second, there are nonperturbative terms in $1-x$ of the form $e^{-\frac{2\pi\,n}{1-x}}\cos(2\pi n y)$ (for $i=1,2,4,5$) and $e^{-\frac{2\pi\,n}{1-x}}\sin(2\pi n y)$ (for $i=3$). The coefficients $\delta_i$ are independent of $y$ and will later parametrise the energy and tension of the non-uniform string.

Requiring the DeTurck vector to vanish then demands
\begin{equation}
\delta _1+7 \delta _2+\delta _4=0
\label{eq:appde}
\end{equation}
which we can use to test the accurary of our numeris.

The above expansion motivates the following field redefinitions.
\begin{equation}
Q_i=1+(1-x^2)^7\widehat{Q}_i\quad\text{for}\quad i=1,2,4,5\,, \qquad\qquad Q_3=(1-x^2)^6\widehat{Q}_3\,,
\end{equation}
so that the boundary conditions near $x=1$ are now replaced by
\begin{equation}
\left.\frac{\partial \widehat{Q}_i}{\partial x}\right|_{x=1}=0
\end{equation}
Note that with the above field redefinitions we can access all the constants $\delta_i$ with at most two derivatives with respect to $x$. In fact, to compute $\delta_i$ with $1,2,3,4$ we don't need to take any derivative at all.

 To compute the energy $\mathcal{E}^{(0)}/G_{11}$ and tension $\mathcal{T}_Z^{(0)}/G_{11}$ densities we follow \cite{Harmark:2003yz,Dias:2007hg,Kraus:1999di} and find
 \begin{equation}
 \mathcal{E}^{(0)}=\frac{\pi ^3 y_+^7}{6720}\left(1024-8 \delta _1-\delta _4\right)\qquad\text{and}\qquad \mathcal{T}_Z^{(0)}=\frac{\pi ^3 y_+^7 }{6720}\left(128-\delta _1-8 \delta _4\right)\,,
 \end{equation}
 where we used Eq.~(\ref{eq:appde}) to replace any occurence of $\delta_2$. Note that in the remaining four patches, we still use $Q_i$
 ; it is only in the two patches nearer to  $x=1$ that we replace $Q_i$ with $\hat{Q}_i$.

 Finally, the entropy density $\mathcal{S}^{(0)}/G_{11}$ and temperature $\mathcal{T}_H^{(0)}/L$ are given by
 \begin{equation}
 \mathcal{S}_H^{(0)}=\frac{16\pi^4 y_+^8}{105} \int_0^{1/2}\sqrt{Q_4(0,y)}\,Q_5(0,y)^4{\rm d}y\qquad\text{and}\qquad \mathcal{T}_H^{(0)}=\frac{7}{4 \pi  y_+}\,,
 \end{equation}
 respectively.
\end{document}